\documentclass[doublecol]{eplArxiv}
\usepackage[switch]{lineno}

\usepackage{graphicx}
\usepackage{amsmath}
\usepackage{verbatim}
\usepackage{cite}
\usepackage[USenglish]{babel}
\usepackage{mathtools}
\DeclarePairedDelimiter{\abs}{\lvert}{\rvert}
\usepackage{subfig}
\usepackage [autostyle, english = american]{csquotes}
\MakeOuterQuote{"}

\pdfoutput=1
\title{Active particle dynamics beyond the jamming density}
\shorttitle{Active particle dynamics beyond the jamming density} 

\author{Daniel R. McCusker\inst{1}, Ruben van Drongelen\inst{1}, \and Timon Idema\inst{1,2}}
\shortauthor{D. McCusker, R. van Drongelen and T. Idema}

\institute{
  \inst{1}Department of Bionanoscience, Kavli Institute of Nanoscience, Delft University of Technology - van der Maasweg~9, Delft, The Netherlands.\\
  \inst{2}\textnormal{Contact:} \normalfont{\email{t.idema@tudelft.nl}}
}


\abstract{Many biological systems form colonies at high density. Passive granular systems will be jammed at such densities, yet for the survival of biological systems it is crucial that they are dynamic. We construct a phase diagram for a system of active particles interacting via Vicsek alignment, and vary the density, self-propulsion force, and orientational noise. We find that the system exhibits four different phases, characterized by transitions in the effective diffusion constant and in the orientational order parameter. Our simulations show that there exists an optimal noise such that particles require a minimal force to unjam, allowing for rearrangements.}

\begin{document}
\maketitle

\section{Introduction}
Many biological systems consist of collections of individuals. Examples include herds of mammals, flocks of birds, bacterial colonies, and tissues. Such systems are intrinsically nonequilibrium, as each microscopic unit composing the system consumes energy to propel itself. One of the earliest self-propelled particle models was proposed by Vicsek \textit{et al.} and describes the flocking behavior of birds. {The original Vicsek system featured point-like self-propelled particles and displayed a second-order phase transition with spontaneous ordering below a critical point \cite{Vicsek1995}. The existence of the ordered phase was apparently in violation of the Mermin-Wagner theorem, which excludes ordered phases for a continuous $O_2$ symmetry in $d=2$. Toner and Tu explained the existence of the ordered phase, attributing it to enhanced density fluctuations as a means of long-range information transfer. Unlike the 2D XY model, which undergoes a Kosterlitz-Thouless transition with only quasi-long ranged order below the critical point, the 2D Vicsek model displays true long-range order \cite{Kosterlitz1973, Toner1995, Toner1998}. A later finite-size scaling analysis showed the actual Vicsek transition to be a nonequilibrium first-order transition, and much work has been done since the original Vicsek model to understand the role of enhanced density fluctuations in nonequilibrium systems \cite{Chate2006, Chate2008, Narayan2007}. Moreover,} variants of self-propelled particle models have emerged {since the Vicsek model in order to capture phenomena like adhesion forces and nematic ordering} ~\cite{Vicsek2012}. In some cases, like flocks of birds, the agents can be adequately modeled as point particles with no excluded volume. For bacterial colonies and tissues, however, the density is so high that the cells are in physical contact with each other. At such a high density, these systems risk becoming jammed. Nonetheless, the constituents rely heavily on rearrangements for their survival, leading us to ask how such colonies prevent jamming.

The phenomenon of jamming has been studied as a feature of granular systems, which are composed of finite-sized particles with purely repulsive potentials. Such systems undergo a phase transition from a "liquidlike" phase to a "solidlike" phase with increasing density, characterized by a sudden arrest of the motion of the constituent particles, which are then locked into a small subset of their phase space. This transition has a nonequilibrium character due to the dissipative forces of particle interactions, yet much work has been done to extend the methods of equilibrium statistical physics to such systems \cite{Behringer2008,Bi2014}. The jamming transition is qualitatively sketched in jamming phase diagrams, which indicate the locations of the jammed and unjammed (solid and liquid) states in temperature-density-stress space; a system becomes jammed at a "high enough" density, with "low enough" stress and temperature \cite{Liu1998, Trappe2001}. For athermal particles in two dimensions with no applied stress, this transition occurs at a critical packing fraction $\rho_c = 0.842$ \cite{OHern2003}.

Previous research has been done to understand the behavior of biological systems at very high density. For instance, in a confluent tissue model with a self-propulsion force, the jammed state is distinguished by a sudden drop in the diffusion constant, as well as a simultaneous change in the cell geometry \cite{Bi2016}; this transition was also observed in experiments \cite{Angelini2011}. Other approaches include vertex models \cite{Alt2017} and Voronoi models \cite{Sussman2017} for epithelial cells. All these self-propelled particle (SPP) models assign a self-propulsion force to each cell, and cover various cell shapes, including point-like, circular disks, elliptical disks, and rods. There are a variety of approaches to the rotational dynamics of SPP models, including self-propelled disks without neighbor alignment \cite{Fily2014}, alignment with the instantaneous velocity \cite{Henkes2011, Sknepnek2015} and torque exerted on nearest neighbors \cite{vanDrongelen2015}. 

So far, the studies that use Vicsek alignment have focused on the rich dynamics. In this letter we consider the arrest of such dynamics as the system approaches jamming. {We sketch a phase diagram in order to relate the jamming transition and the ordering transition, and we show that the system is able to unjam both below and above the critical noise at densities well above the passive jamming density. We quantify the amount of rearrangement in the system by introducing an effective diffusion constant and by studying the scaling of density fluctuations. We measure the system's velocity correlation function above and below the transition in order to understand the effect of interparticle interactions on the system's dynamics near jamming. We find an optimal noise such that particles require a minimal self-propulsion force to unjam, and we also find, surprisingly, that the system is able to order even in the jammed state, in the absence of density fluctuations.}

\section{Model system}
\label{sec:model}
We place $N$ soft, self-propelled particles in a square with linear size $L$ and periodic boundary conditions, and impose the dimensionless packing fraction $\rho = \sum_{i=1}^N \pi a_i^2 / L^2$. To prevent crystallization, the radii of the particles $a_i$ are drawn from a Gaussian distribution with mean $\mu = \bar{a}$ and standard deviation $\sigma = \bar{a}/10$. We consider systems in which viscous forces dominate over inertial forces, e.g. cell tissues or colonies of unicellular organisms. The dynamics of the particles are then overdamped and governed by Stokes' law:
\begin{equation}
  \vec{F}_i = \zeta_i \vec{v}_i.
  \label{eq:Stokes}
\end{equation}
In eq. 1, particle $i$ moves at velocity $\vec{v}_i$ in proportion to the total force $\vec{F}_i$ exerted on it. The proportionality constant $\zeta_i$ depends on the viscosity $\eta$ and the particle's radius $a_i$, and is given by $\zeta_i = (32/3) \eta a_i$ in two dimensions and $\zeta_i = 6 \pi \eta a_i$ in three dimensions \cite{Landau1987}. The total force on the particle is the sum of steric repulsion forces with all particles $j$ that generate overlap, $\vec{F}_{\text{rep}, j}$, and a self-propulsion force, $\vec{F}_\text{sp}$. We choose a simple harmonic repulsion, such that the force is proportional to, and in the direction of, the linear overlap $\vec{d}_{ij}$. Hence, the total force is
\begin{align}
    \vec{F}_i &= \sum_j \vec{F}_\text{rep} + \vec{F}_\text{sp} = \sum_j k \vec{d}_{ij} + F_\text{sp}\hat{\theta}_i \\
    &= \sum_j  \vec{d}_{ij} + \lambda_s  a_i \hat{\theta}_i,
    \label{eq:EoM}
\end{align}
where $k$ is the spring constant for the repulsion force. In eq.~\ref{eq:EoM}, we scale all distances in units of the average particle radius $\bar{a} = 1$ and choose $k=1$, setting the characteristic force $k \bar{a} = 1$. This allows us to define the dimensionless parameter $\lambda_s = F_\text{sp}/k\bar{a}$ which sets the strength of the self-propulsion in our simulations. The self-propulsion term is proportional to the particle's radius so that all particles would move at the same velocity in the absence of overlaps.

The unit vector $\hat{\theta}_i$ indicates the direction of each particle's self-propulsion, which is determined by the Viscek alignment rule. At each time step, each particle "senses" the orientations of its neighbors, and aligns itself with its neighbors' average direction. This sensing occurs with some error, which we model as a noise $\Delta\theta$. We can then write the for orientation of particle $i$: 
\begin{equation}
  \hat{\theta}_i = \langle\hat{\theta}_j\rangle_{j \in \mathcal{N}_i} + \Delta\theta,
  \label{eq:orientation}
\end{equation}
where $\mathcal{N}_i$ is the set of particles in the neighborhood of particle $i$. We define the neighborhood to be the region of space enclosed by a radius of $2.8 \bar{a}$ around the center of particle $i$. We choose this distance such that two neighboring large particles are considered neighbors, while two small particles separated by a third are not.
The random rotation angle $\Delta\theta$ is drawn from a uniform distribution on $\lambda_n\big[{-\pi}, \pi\big]$. $\lambda_n = 0$ represents total alignment, while for $\lambda_n = 1$, each particle performs a random walk with no alignment with its neighbors.
The order parameter quantifies the global alignment in the system, and is defined as
\begin{equation}
  \phi = \frac{1}{N} \left| \displaystyle \sum_{i=1}^{N} \frac{\vec{v}_i}{\abs{\vec{v_i}}} \right|,
\label{eq:orderParameter}
\end{equation}
which equals unity when all particles move in parallel, and approaches zero when the system has no global order.

To define our time scale, we realize that two average-sized overlapping particles in an overdamped regime will see their overlap decrease according to an exponential with a decay time $\tau_{\text{relax}} = \bar{\zeta}/k$. We choose this physical time to be the unit time, $\tau_{\text{relax}} = 1$. We also note that there is a time interval at which the particles orient themselves, $\tau_{\text{orient}}$. We fix a timestep $\Delta t = {\tau_{\text{orient}}}/{\tau_{\text{relax}}} = 0.1$ so that there are several "sensing events" for each relaxation time. With this choice of scaling, we can write our equations of motion as:
\begin{align}
\label{eq:velocityrule}
    \vec{v}_i &= \sum_j\frac{\vec{d}_{ij}}{a_i}  + \lambda_s \hat{\theta}_i, \\
\label{eq:orientationrule}
    \hat{\theta}_i &= \langle\hat{\theta}_j\rangle_{j \in \mathcal{N}_i} + \Delta\theta,
\end{align} 
which we integrate at every time step according to 
\begin{equation}
    \vec{x}_i(t+\Delta t) = \vec{x}_i(t) + \vec{v}_i(t)\Delta t. 
\end{equation}
Our system now has three free parameters: the packing fraction $\rho$, self-propulsion force $\lambda_s$, and noise $\lambda_n$. The time step $\Delta t$ is not a free parameter; it is coupled to the self-propulsion velocity, and so it sets the scale of $\lambda_s$. In our simulations, we allow the system to relax for $1 \times 10^5$ time steps, during which we let the particles rearrange as passive particles and come to a stable configuration. We then thermalize the system for an additional $1 \times 10^6$ time steps, during which we slowly and linearly ramp up the self-propulsion force to its final value, at which point we begin measurements. 

\section{Results and Discussion}
The jamming transition for soft, passive particles is at $\rho_c = 0.842$; we ran simulations for densities $\rho \in \{0.88$, $0.94$, $1.00\}$. {Except where noted, the simulation results are for systems of $N = 1\times 10^5$ particles.}

\subsection{Simulation screenshots}

\begin{figure}

  \subfloat[]{\includegraphics[width=0.3\columnwidth]{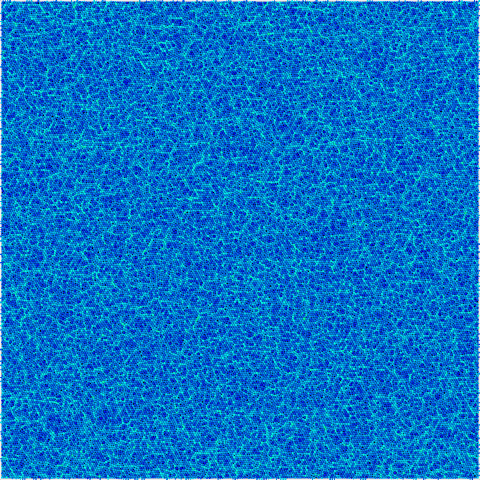}}
    \hfill
  \subfloat[]{\includegraphics[width=0.3\columnwidth]{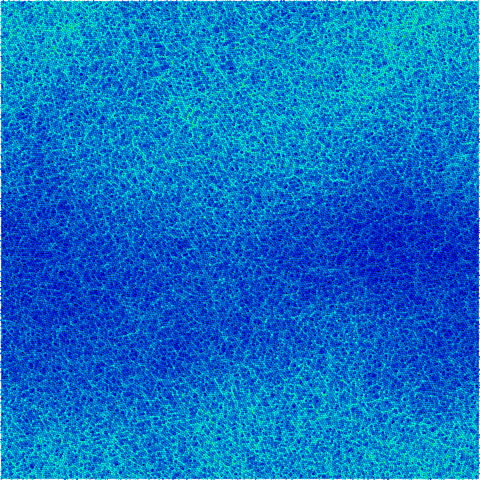}}
    \hfill 
  \subfloat[]{\includegraphics[width=0.3\columnwidth]{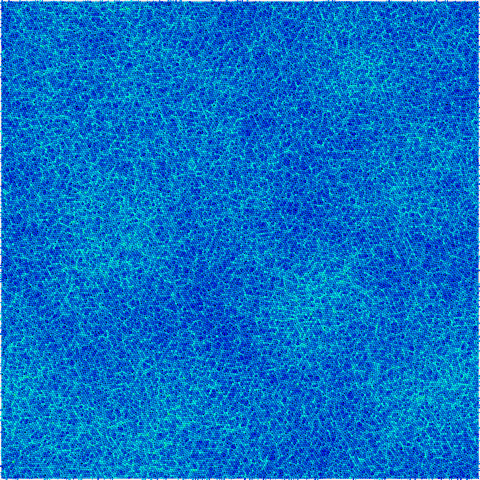}}
        
\caption[Simulation screenshots.] {{Screenshots for simulations with $5 \times 10^4$ particles, showing the qualitative phase behavior. Individual particles are shaded to indicate the amount of overlap: dark blue particles have little overlap, and light blue particles have more overlap. In (a) the system is ordered, but jammed, with a low self-propulsion force, and the pressure is homogeneously distributed across the system. In (b) we increase the self-propulsion force and the system unjams, forming a Vicsek-like ordered state ($\phi>0$) with giant density fluctuations, indicated by the existence of a high-density region and a low-density region. This gives the particles plenty of space to rearrange, and is accompanied by an increase in the effective diffusion constant. In (c), we increase the noise above the critical noise so that the global ordering vanishes ($\phi=0$) and the giant fluctuation disappears, replaced by smaller fluctuations of lesser magnitude. The effective diffusion constant drops, but smaller "flocks" form and the system is still unjammed. The simulations are placed in the phase diagram of fig. \ref{fig:MSDphaseFull}}.} 
\label{fig:screenshots}
\end{figure}

{In fig. \ref{fig:screenshots} and in the supplementary videos, we show simulations at three different conditions, resulting in three qualitatively different states. \ref{fig:screenshots}(a) illustrates a jammed state, \ref{fig:screenshots}(b) an unjammed state in the ordered phase, and \ref{fig:screenshots}(c) an unjammed state in the disordered phase. These systems are highlighted in the phase diagram (fig. \ref{fig:MSDphaseFull}).}

\subsection{Mean-squared displacement and effective diffusion}

A perfectly ordered system ($\lambda_n = 0 \text{ and } \phi = 1$) will exhibit global translation {in the lab frame}, but no rearrangements, as each particle aligns perfectly with its neighbors. We therefore measure particle trajectories by considering the mean-squared displacement from the center of mass $\bar{x}$ of the system:

\begin{equation}
    \text{MSD}(t) = \frac{1}{N}\sum\limits_{i=1}^{N}\bigg[ \big(x_i(t) - x_i(0)\big) - \big(\bar{x}(t) - \bar{x}(0)\big)\bigg]^2. 
\end{equation}
In a jammed system, particles are not able to travel more than a distance $\bar{a}$ because they are obstructed by their neighbors. This effect is known as caging, and is visible in the mean-squared displacement as a plateau. In order to quantify the rearrangements in our system, we performed a linear fit with the assumption that $\text{MSD}(t) = 4Dt$, which would be the case for a perfect random walk in two dimensions. We ignore the first $1 \times 10^6$ time steps, the transient, in the linear fits.

\begin{figure}
    \centering
        \subfloat[]{\includegraphics[width=0.499\columnwidth]{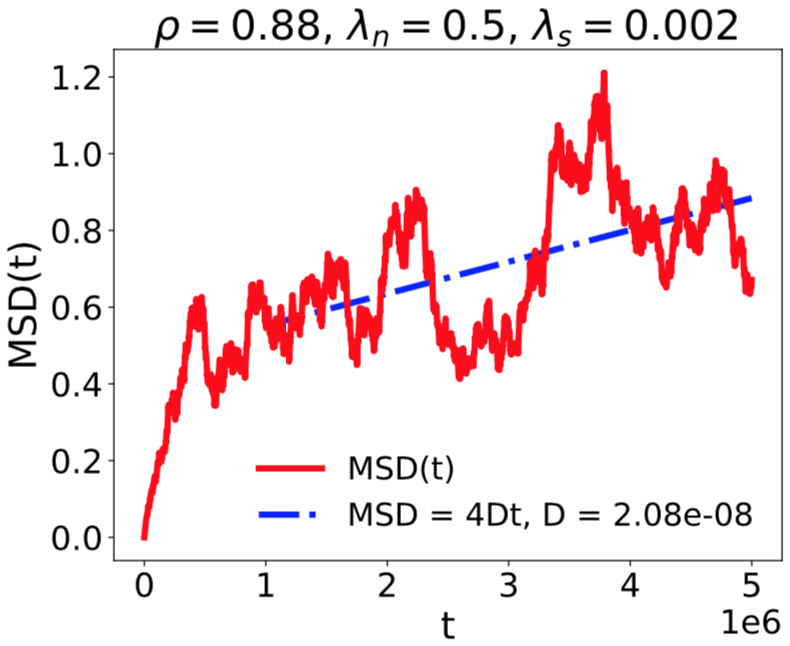}}
        \hfill
        \subfloat[]{\includegraphics[width=0.499\columnwidth]{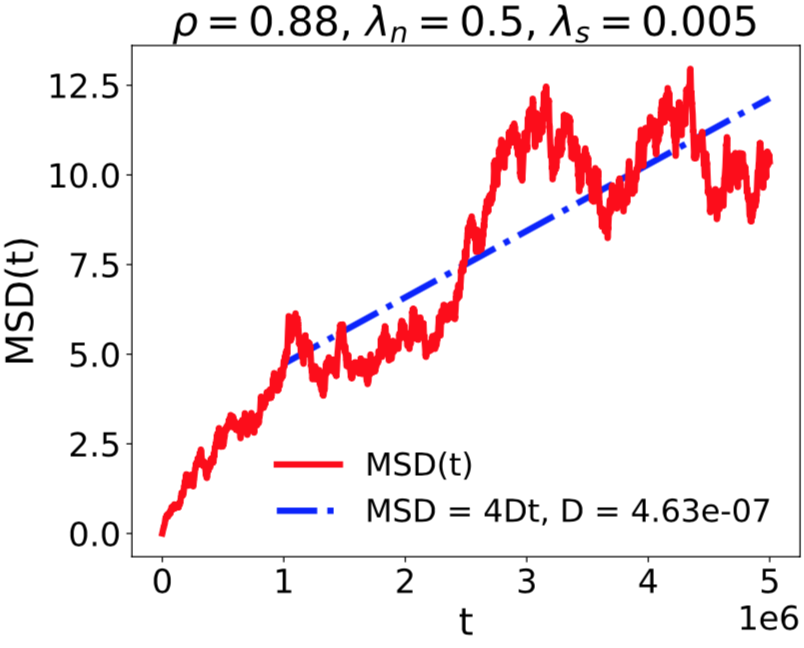}}
        \\
        \subfloat[]{\includegraphics[width=0.499\columnwidth]{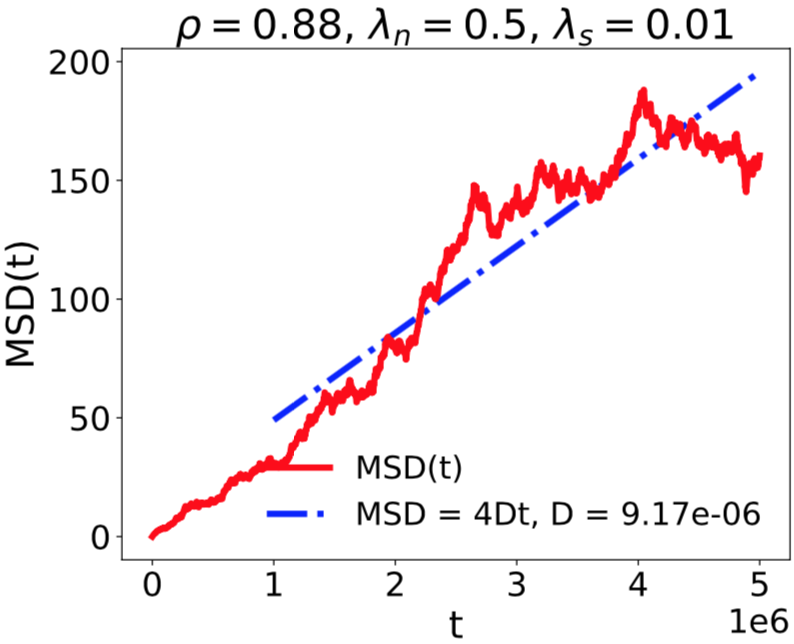}}
        \hfill
        \subfloat[]{\includegraphics[width=0.499\columnwidth]{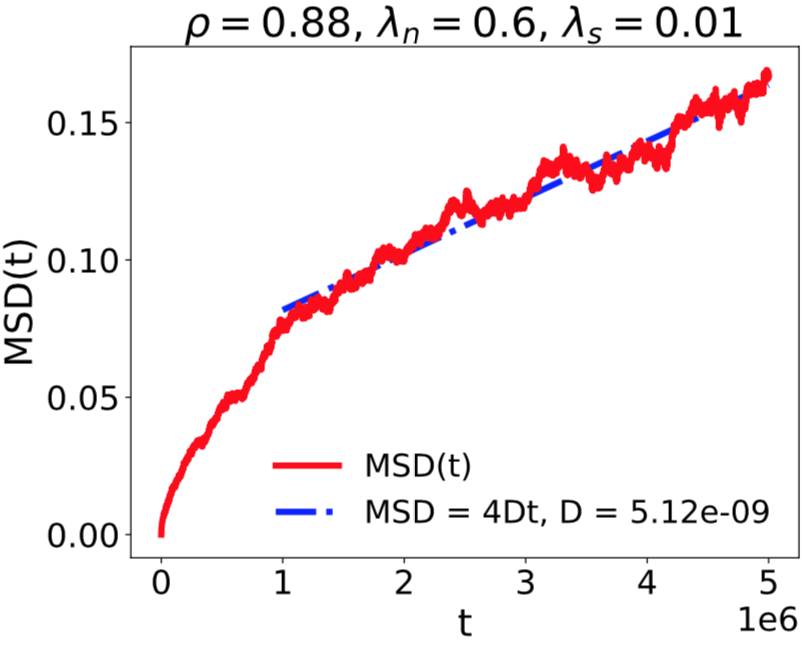}}
   
    \caption{Selected MSD curves for simulations with $1 \times 10^5$ particles at a density of $\rho = 0.88$. Every particle has rearranged {at least} once if the value of the MSD is greater than~1. From (a) through (c), we increase $\lambda_s$ and find that an initially jammed system is able unjam when the particles push harder. In (d), we increase the noise and find that the system jams, even for large self-propulsion speed. 
    {These four curves are highlighted in fig. \ref{fig:MSD}b}.}
    \label{fig:MSDcurves}
\end{figure}

\begin{figure}
\centering
\includegraphics[width=0.8\columnwidth]{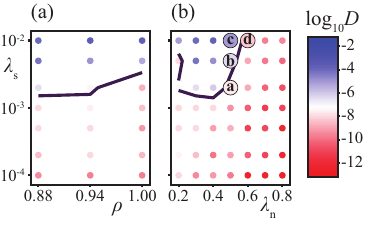}
\caption{Phase diagram slices from mean-squared displacement analysis, showing the values of the calculated diffusion constant in the $\rho$ - $\lambda_s$ - $\lambda_n$ parameter space. The blue dots represent unjammed systems and the red dots represent jammed systems. The interpolated contour roughly indicates the self-propulsion strength necessary for unjamming. In (a), we fix the noise value $\lambda_n = 0.3$ and find that the effect of increasing the density is to increase the self-propulsion force required for unjamming. In (b) we show the effect of noise on the jamming dynamics for constant density $\rho = 0.88$. {The four points indicated \textbf{a}, \textbf{b}, \textbf{c}, \textbf{d} are the four MSD curves in fig. \ref{fig:MSDcurves}}. We know that the system is always jammed for $\lambda_n = 0.0$, and we see that the system also jams for $\lambda_n \geq 0.6$ for the range of $\lambda_s$ we measured here. The valley suggests an optimum noise for unjamming.}
\label{fig:MSD}
\end{figure}

Some examples of MSD curves for both jammed and unjammed states are shown in fig. \ref{fig:MSDcurves}. The curves in fig. \ref{fig:MSDcurves} reveal that small changes in noise or self-propulsion speed can have order-of-magnitude effects on the effective diffusion constant. Therefore, near jamming, the calculated values of $D$ provide an approximation for the amount of rearrangement in the system. We consider our system to be jammed if the MSD plateaus at $\text{MSD}(t) < \bar{a}^2$, or if the fitted diffusion constant does not allow for displacements larger than $\bar{a}$ during our measurement of the MSD. Note that $D$ does not indicate Brownian motion, because particle displacements result from the particles' activity rather than the conversion of heat to kinetic energy. The calculated values of $D$ are shown in fig.~\ref{fig:MSD}. We set the diffusion coefficient for marginally jammed systems to $\left({4\times 4\times10^{6}}\right)^{-1}$. Jammed systems are shown in red in fig.~\ref{fig:MSD}, while unjammed systems are shown in blue. We find unjammed systems even at densities far exceeding the jamming density when the particles exert a sufficiently large self-propulsion force.


\begin{figure}
\centering
\includegraphics[width=0.7\columnwidth]{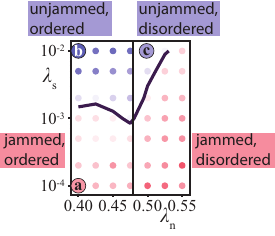}
\caption{{Phase diagram near the critical point for a density $\rho=0.88$, showing the relationship between the ordering transition and the jamming transition. The colors are the same as in fig.~\ref{fig:MSD}. The contour separates jammed from unjammed states, while the vertical line separates the ordered ($\phi > 0$) from the disordered ($\phi = 0$) state. The points denoted \textbf{a}, \textbf{b}, and \textbf{c} are the simulations illustrated in figure 1. An initially jammed system \textbf{(a)} below the critical noise can unjam when the self-propulsion force is increased so that the particles are able to push past their neighbors \textbf{(b)}. Increasing the noise just above the critical point will cause the order parameter to fall to zero, though the system will remain unjammed \textbf{(c)}. Increasing the noise further will jam the system, as the correlation length decreases (see fig.~\ref{fig:allcorrelation}).}}
\label{fig:MSDphaseFull}
\end{figure}

\begin{figure}
\centering
\includegraphics[width=1.0\columnwidth]{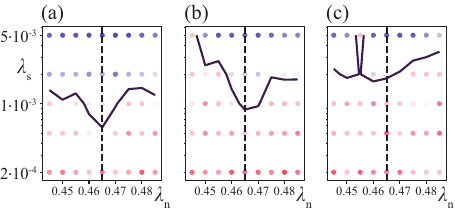}
\caption{Phase diagram near the critical point, for $\rho=0.88$ (a), for $\rho=0.94$ (b), and for $\rho=1.00$ (c). The colors are the same as in fig.~\ref{fig:MSD}. {We use the position of the minimum in each of these three plots to estimate a critical point $\lambda_c = 0.465$ at which transport is enhanced.} {The minimum is less obvious for $\rho = 1.0$, but the value of $D$ is enhanced at $\lambda_n = 0.465$ for $\lambda_s = 0.05$ (uppermost values, dark blue).}}
\label{fig:MSDclose}
\end{figure}

The contour separating jammed from unjammed systems indicates the required self-propulsion strength to unjam for a given noise and density. We see that the effect of increasing the particle density is to increase the required self-propulsion force for unjamming; in more crowded environments, particles must push harder to get past their neighbors. In fig. \ref{fig:MSDphaseFull}, we {zoom in to the contour's minimum}, and find that particles are able to marginally unjam for $\lambda_s = 1 \times 10^{-3}$ and $\lambda_n = 0.475$. {We zoomed in even more in fig. \ref{fig:MSDclose} and found a minimum of $\lambda_n$ at $\lambda_c \approx 0.465$ for three different densities}. 

Time-averaging the MSD curves proves difficult, as the particle trajectories are contingent on their histories. Furthermore, the simulations are already computationally demanding, so averaging over many simulations {or longer times} is infeasible. We therefore will study other quantities to give further insight into the dynamics near jamming. 

\subsection{Order-disorder phase transition}
\begin{figure}
\centering
\includegraphics{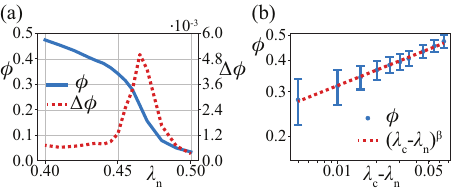}
\caption[Dependence of the order parameter $\phi$ on $\lambda_n$.] {Dependence of the order parameter $\phi$ on $\lambda_n$ at $\rho = 0.88$ and $\lambda_s = 0.002$. The data in (a) suggests a critical point near $\lambda_n = 0.465$, where the order parameter falls to zero and the fluctuations peak, {though the smoothness of the curve near the critical point suggests persistent finite-size effects}. Fig.~(b) shows $\phi$ on a log-log scale, with a power-law fit: $\phi \sim (\lambda_c-\lambda_n)^\beta$, with $\lambda_c = 0.465 $ and $\beta = 0.206 \pm 0.004 $.} 
\label{fig:orderNoise}
\end{figure}

{Motivated by the existence of long-range order in the Vicsek model and its relevance for active jamming}, we studied the behavior of the order parameter in our system of finite-sized particles, and found an order-disorder transition near a critical noise value of $\lambda_n = \lambda_c = 0.465$, as shown in fig. \ref{fig:orderNoise}. Figure~\ref{fig:MSDclose} shows higher values of the effective diffusion constant near this critical noise value, indicating that transport is enhanced near the critical point. We speculate that biological systems might sit near this critical value so that they are able to rearrange with a minimum of effort. Furthermore, {when we combine the two phase transitions in a single plot, fig.~\ref{fig:MSDphaseFull}, we find an additional unjammed state which is disordered, indicating that Vicsek-like flocking is not necessary for unjamming. This phase is also illustrated in the screenshot in fig.~\ref{fig:screenshots}c and in the curves in figs.~\ref{fig:MSDcurves}b-c.}

Below the critical noise, our system shows spontaneous ordering, consistent with a second-order phase transition at the critical point $\lambda_c$. Furthermore, the variance in the order parameter, $\sigma^2 = \langle \phi^2 \rangle - \langle \phi \rangle^2$ shows a sharp peak at the transition, a characteristic of second-order transitions (fig.~\ref{fig:orderNoise}a). {This result is different from both the XY model (no ordering in the thermodynamic limit) and the Vicsek model (first-order phase transition in the thermodynamic limit).} In fig.~\ref{fig:orderNoise}b, we plot the value of the order parameter $\phi$ as a function of $\abs{\lambda_c - \lambda_n}$ on a log-log scale near the critical point. We find power-law behavior, with a critical exponent $\beta$. The uncertainty in $\beta$ comes from the reported confidence of the fit, while the error bars indicate the standard deviation in the order parameter for a given noise value, $\sigma = \sqrt{\langle \phi^2 \rangle - \langle \phi \rangle^2}$. 

\subsection{Relationship between the jamming transition and the order-disorder transition}

{Fig. \ref{fig:MSDphaseFull} shows that the jamming and ordering transitions result in four distinct phases. The unjammed, ordered phase is characterized by Vicsek-like dynamics with enhanced density fluctuations. Increasing the noise results in a disordered phase where the system is unjammed, characterized by small coherent flocks which collide, allowing for particle rearrangements at their boundaries (note that "disordered" here refers to fact that $\phi=0$ and is not a statement about the packing geometry). Decreasing the self-propulsion strength or increasing the noise further from this phase moves the system to a state which is jammed and disordered. Surprisingly, decreasing the noise at low self-propulsion speed can result in a jammed, ordered state, with suppressed density fluctuations, which we seek to explain in the following section.}

\subsection{Long-range order and ordering in the jammed state}

\begin{figure}
\centering
\includegraphics[scale=1]{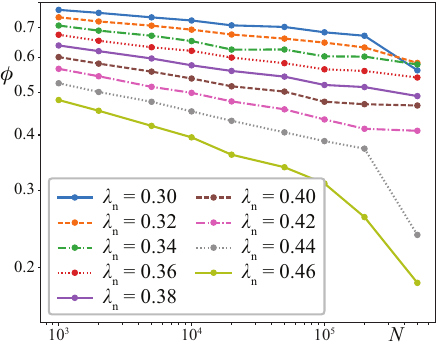}
\caption{Finite-size effects: how the order parameter depends on system size in the ordered, jammed phase. Each curve represents one fixed $\lambda_n \in [0.30,\,0.46]$. All curves are for $\rho = 0.94$ {and a small self-propulsion force $\lambda_s=1\times 10^{-4}$}. We averaged the order parameter for only $1\times 10^5$ time steps in order to allow the study of large systems. For a KT transition which obeys the Mermin-Wagner theorem, the order parameter should fall to zero in the infinite size limit.}
\label{fig:orderScaling}
\end{figure}

{Fig. \ref{fig:orderNoise} showed that long-range order can exist in the unjammed phase. Because the particles interact via Vicsek alignment, we expected that the system might display long-range order, in accordance with the Toner-Tu theory \cite{Toner1995, Toner1998}. To investigate if true long-range order exists in the jammed state, we measured the order parameter as a function of system size, up to a maximum system size of $5\times 10^5$ $\left(700 \times 700 \right)$ particles (fig.~\ref{fig:orderScaling}). Our results suggest that the order parameter for a jammed system barely falls for system sizes varying over three orders of magnitude. If the jammed state were to display a Kosterlitz-Thouless (KT) transition at the critical noise, the ordered phase should not exist in the thermodynamic limit, in agreement with the Mermin-Wagner theorem. Density fluctuations are suppressed in the jammed state, and so can not be an ordering mechanism as in the Toner-Tu theory (see also fig. \ref{fig:densityFluct}). We might expect the Mermin-Wagner theorem to take over in this phase and destroy long-range order.}

{We considered several possible mechanisms for long-range information transfer. Each particle in this model has an average of six nearest neighbors, more than in the prototypical XY model, so it could be that any "spin waves" forming in this system are "stiffer" than in the XY model and so have a longer wavelength, pushing the thermodynamic limit to larger sizes. Moreover, the geometry of the packing (roughly hexagonal with defects) allows for more paths between a given particle pair than in the XY model, as well as shorter diagonal paths, in contrast to the strictly "Manhattan" distances of the XY lattice. The geometry could therefore provide a mechanism for enhanced information transfer. A packed colony of cells could possibly exploit this geometry in order to "communicate" across a system which is large in terms of biological length scales, even though strictly speaking, the system would be disordered in the limit of infinite size. Henkes \textit{et al.} also proposed a mechanism of information transfer due to an active, jammed system's vibrational modes \cite{Henkes2011}; this mechanism may be at work here as well.}

\subsection{Correlation function}

\begin{figure}
\centering
\includegraphics[width=1.0\columnwidth]{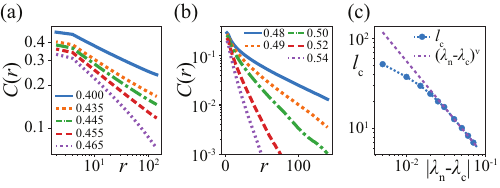}
\caption[Velocity correlation function.] {The velocity correlation function in the ordered (a) and disordered (b) phases, {where the legend indicates the value of $\lambda_n$}. All displayed curves are for $\rho = 0.94$ and $\lambda_s = 1\times 10^{-3}$. In both (a) and (b), the nearest-neighbor correlation appears lower than expected because overlapping particles are repelling each other. We therefore exclude this value from our fits. The curves begin to cross over to exponential decay around $\lambda_c = 0.465$. The fitted values with increasing $\lambda_n$ are $\eta = 0.26, 0.31, 0.33, 0.33, 0.27$. In (b), the value of the correlation length falls; with increasing noise, $l_c = 37.4, 24.3, 17.3, 10.2, 7.0, 3.5$. In (c), we plot the fitted correlation lengths as we approach the critical noise from above, and fit the six rightmost points to a power law $l_c \propto \abs{\lambda_n - \lambda_c}^{\nu}$. The fit is not as good near the critical noise; we can also see this in (b) near the critical noise, where the fits begin to deviate from straight lines. Notably, just above the transition, the correlation length extends over a significant number of particles, suggesting the formation of small, coherent flocks while in the globally disordered phase. {This result differs from the case of the 2D XY model, in which the exponent $\nu$ is not defined; instead, the correlation length diverges even faster than a power-law, according to $l_c \propto  \exp\left(b t^{-1/2}\right)$, with $t$ the reduced temperature and $b$ a constant \cite{Kosterlitz1974}}.} 
\label{fig:allcorrelation}
\end{figure}

To better understand the {unjammed state in the disordered phase}, we measured the velocity correlation

\begin{equation}
    C(r) = \frac{1}{n(r)}\sum_{\{i,j \mid r_{ij}=r\}}\frac{\vec{v_i}\cdot\vec{v_j}}{\abs{\vec{v_i}}\abs{\vec{v_j}}} ,
    \label{eq:correlation}
\end{equation}
where $n(r)$ is the number of pairs separated by a distance $r$. We find that the form of the correlation function is a power-law decay below the critical noise, and changes to an exponential decay above the critical noise, with a correlation length which decreases with increasing $\abs{\lambda_n~-~\lambda_c}$, shown in fig.~\ref{fig:allcorrelation}:

\begin{equation*}
    C(r) \propto 
    \begin{cases} 
        \frac{1}{r^{2-d+\eta}} & \lambda_n < \lambda_c, \\
        e^{-r/l_c} & \lambda_n > \lambda_c. \\
    \end{cases}
\end{equation*}
    
Though the system is globally disordered for noise values above $\lambda_c$, a non-vanishing correlation length indicates that small, coherent flocks can still form in the disordered phase. This suggests a mechanism for unjamming in which flocks collide and allow rearrangements at their boundaries. 

\subsection{Density fluctuations}

We can further quantify the system's dynamics just above the critical point by considering the scaling of the density fluctuations in the system. In the grand canonical ensemble of equilibrium statistical physics, point particles enter and leave a system with independent probabilities, and the number fluctuations scale as 

\begin{equation}
    \Delta N \propto N^{1/2} . 
\end{equation}

It is known that active systems exhibit "giant number fluctuations", reflected in a scaling exponent $m > \frac{1}{2}$, due to orientational coupling\cite{Chate2006, Narayan2007}. With our finite-sized particles, we measure area fluctuations in a "measurement circle" of a given size, centered at the system's center of mass, and fit the scaling exponent $m$:

\begin{equation}
    \Delta A \propto A^{m} . 
\end{equation}

\begin{figure}
\centering
\includegraphics[width=0.9\columnwidth]{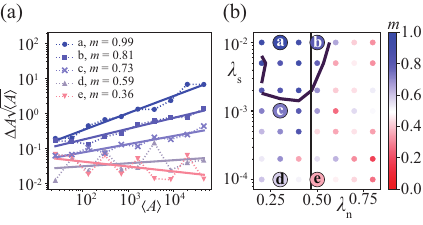}
\caption{{The scaling of density fluctuations $\Delta A$ with expected area $\langle A \rangle$; the scaling follows the relationship $\Delta A~\propto~\langle A \rangle^m$. All curves are for constant density $\rho = 0.88$. A horizontal line in (a) indicates a scaling exponent $m=\frac{1}{2}$. In (a) we show the results of five simulations, and in (b) we draw the phase boundaries from fig. \ref{fig:MSD} and place the simulations from (a) in the parameter space. \textbf{a}: in the ordered, unjammed phase, the system shows Vicsek-like density fluctuations with $m\approx 1$. \textbf{b}: in the disordered, unjammed phase, particles are rearranging, but the global density wave in the Vicsek phase is suppressed in favor of smaller flocks. \textbf{c}: In the jammed, ordered state, just below the jamming transition, the particles are pushing on each other and forming small density waves, though not quite hard enough to unjam. \textbf{d}, \textbf{e}: in the jammed state with very low-self propulsion force, the actual values of the fluctuations $\Delta A$ are very small, resulting in poorer linear fits.}} 
\label{fig:densityFluct}
\end{figure}

{Figure \ref{fig:densityFluct} shows the values of $m$ in the $\lambda_n$-$\lambda_s$ parameter space. The value of $m$ changes smoothly across the phase boundaries, but gives insight into the dynamics in the different phases. In the ordered, unjammed phase, $m\approx1$, in agreement with results for point particles, and as illustrated in fig. \ref{fig:screenshots}b \cite{Chate2006, Narayan2007}. When the noise increases, the system becomes disordered, though small flocks form and the system is still unjammed, as illustrated in fig. \ref{fig:screenshots}c. A jammed system in the ordered state, just below the jamming transition, can show weak scaling, indicating that the particles are forming small density waves, but are not pushing hard enough to unjam. When the self-propulsion force decreases even further, the density fluctuations are totally suppressed, as in figure \ref{fig:screenshots}a.}

\section{Conclusions}
We simulated a collection of soft, self-propelled particles with Vicsek alignment at packing fractions $\rho \in \{0.88$, $0.94$, $1.00\}$, well above the passive critical packing fraction $\rho_c = 0.842$. We constructed a phase diagram in the $\rho$~-~$\lambda_s$~-~$\lambda_n$ space which distinguishes jammed and unjammed packings. Systems jam for too little noise, as the particles align in parallel and cannot push aside their neighbors. They also jam for too much noise, in which case the particle velocities decorrelate from those of their neighbors, and the particles become caged.

The system undergoes a second-order phase transition; below a critical noise $\lambda_n = \lambda_c$ the system spontaneously orders. This is indicated by power-law behavior of the order parameter near the critical point, and by the form of the correlation function, which crosses over from exponential to power-law decay. {This behavior differs from the point-like Vicsek model, which shows a first-order transition, and the XY model, which does not order in $d=2$}. 

With a strong enough self-propulsion force, particles can unjam in both the ordered and disordered phases. In the disordered phase, particles are able to unjam by forming small, colliding flocks. By fitting the correlation function, we found a nonvanishing correlation length for these systems, with a noise value $\lambda_n > \lambda_c$. The correlation length roughly follows a power-law divergence just above $\lambda_c$, {again in contrast to the XY model}.

There is an optimal noise value near $\lambda_c$ at which particle transport is enhanced, and particles are able to unjam with a minimum self-propulsion. The exact value of the minimum is difficult to determine due to the noisy dynamics, illustrated in the MSD curves. A system of $5\times 10^5$ ($700 \times 700$) particles shows surprising ordering even in the jammed state (very small $\lambda_s$). Even though the particles are barely moving, the information about their velocities is able to propagate across the entire system. This happens without the enhanced density fluctuations displayed by the active, unjammed systems, and could result from the {vibrational modes or} packing geometry. 

\acknowledgments
This work was supported by a grant from the Fulbright U.S. Student Program and the TU Delft Faculty of Applied Sciences. It was also supported by the Netherlands Organization for Scientific Research (NWO/OCW), as part of the Frontiers of Nanoscience program.

\end{document}